\documentclass{aastex}
\usepackage{spr-astr-addons}
\usepackage{url}

\RequirePackage{color}

\begin{document}

\title{The Pioneer anomaly and the holographic scenario}
\slugcomment{Not to appear in Nonlearned J., 45.}
\shorttitle{The Pioneer anomaly}
\shortauthors{Jaume Gin\'e}

\author{Jaume Gin\'e\altaffilmark{1}}
\affil{Departament de Matem\`atica, Universitat de Lleida,\\
Av. Jaume II, 69. 25001 Lleida, Spain}


\begin{abstract}
In this paper we discuss the recently obtained relation between the Verlinde's
holographic model and the first phenomenological Modified Newtonian dynamics.
This gives also a promising possible explanation to the Pioneer anomaly.
\end{abstract}



\section{The phenomenological version of MOND}

The Modified Newtonian dynamics theory (MOND) was introduced by
Milgrom to solve the galaxies rotation curves problem as an alternative to the dark matter.
The MOND can be implemented by a modification of the Newton's second law
or the Newton's law of gravity.

In particular Milgrom (in the formulation where the Newton's second law is modified)
allowed for an inertia term not to be proportional to the
acceleration of the object but rather to be a more general
function of it. More precisely, it has the form
\[
m_i \ \mu (a/a_0) \ {\bf a} = {\bf F},
\]
where $\mu(x \gg 1)\approx 1$, and $\mu(x \ll 1)\approx x$ and
$a=|{\bf a}|$, replacing the classical form $m_i \ {\bf a} = {\bf
F}$. Here $m_i$ is also the inertial mass of a body moving in an
arbitrary static force field ${\bf F}$ with acceleration ${\bf
a}$, see \cite{Mi1}. For accelerations much larger than the
acceleration constant $a_0$, we have $\mu \approx 1$, and
Newtonian dynamics is restored. However for small accelerations
$a \ll a_0$ we have that $\mu=a/a_0$. In this case if ${\bf F}$ is the
gravitational force of a central mass $M$, then the modulus of the acceleration
is $a= \sqrt{a_0 G M}/r$. This acceleration gives a constant velocity
$v= \sqrt[4]{GMa_0}$ in a circular orbit and the correct value of the
galactic rotational curves. However, it has been shown that
Milgrom theory, while solving a few difficulties, gives rises
to other fresh problems, see for instance \cite{F,S}. The fundamental objection to a modification of
the inertia is that it violates the equivalence principle, tested
to an accuracy of $10^{-13} \ Kg$, see \cite{B}, and the energy
conservation. The version of MOND presented is not a consistent
theory and it is only a phenomenological approach. To solve these
problems Bekenstein and Milgrom proposed in \cite{BM} a
nonrelativistic potencial theory for gravity which differs from
the Newtonian one. In \cite{FB} simple analytical forms of $\mu(x)$ are analyzed and
satisfactory fits to the observationally determined terminal velocity curve are obtained. A
theoretical argument that supports a certain form of $\mu(x)$ against other is still not known.
In \cite{G} we made a first approximation to the problem and deduced the following form of $\mu(x)$,
in the context of the Mach's principle
\begin{equation}\label{eq}
m_i \left (\frac{a}{a + a_0}\right){\bf a }= {\bf F}.
\end{equation}
This simple form of $\mu(x)$ yields
very good results in fitting the terminal velocity curve of Milky
Way and others, see \cite{FB}. Moreover, in \cite{G2} a new form for the $\mu(x)$
appearing in the Milgrom formula was obtained:
\begin{equation}\label{for}
m_i \left (\frac{|{\bf a}|}{|{\bf a} + {\bf a_e}|}\right){\bf a }=
{\bf F},
\end{equation}
where ${\bf a_e}$ is an effective acceleration given by ${\bf
a_e}={\bf a_0} (1 - R_{obs}/R_U) {\bf \hat{R}_U}$, $R_{obs}$ is
the distance to the object and $R_U$ is the radius of the
causal connected universe. For local objects we have $a_e \sim a_0$ and for far away
objects $a_e \sim 0$. Equation (\ref{for}) contains (\ref{eq}) as a particular case.
The form of $\mu(x)$ presented in \cite{G2} is a modification of the inertia following
the ideas developed by Milgrom in \cite{Mi5,Mi3}, using the relativity principle of motion
and assuming the proven fact of the accelerated
expansion of the universe. In the formula (\ref{for}) we have a
vectorial sum of accelerations and depending if the vectors are
quasi-collinear or are perpendicular the vectorial sum gives different
values. In \cite{G2} it is also established a relation
between the MOND and the deceleration parameter of the expansion of the universe.

\section{Verlinde holographic scenario}

Verlinde propose a model where the second Newton law and Newton's law of gravitation arise from
basic thermodynamic mechanisms. In the context of Verline's holographic model, the response of a body to the force may be understood in terms of the first law of thermodynamics. We consider a holographic screen in the plane $yz$
that intersects de $x$ axis at $x+ \Delta x$, where $\Delta x$ is a small increment distance. As the body approaches the screen, its descriptive information becomes encoded holographically on the screen. The entropy of the screen increase by some amount $\Delta S$. In a similar way in which a particle approaching the event horizon of a Schwarzschild black hole increases the entropy of the horizon, in \cite{V} is proposed that
\begin{equation} \label{1}
\Delta S = 2 \pi k_B \frac{m c}{\hbar} \, \Delta x.
\end{equation}
When the body traverse the distance $\Delta x$, its energy changes by an amount $\Delta E = F \Delta x$, which is the incremental work done by the force $F$. Using the first law of thermodynamics, the model sets that
\begin{equation} \label{2}
F \Delta x = T \Delta S.
\end{equation}
An observer in an accelerated frame experiences the associated Unruh \cite{U} temperature
\begin{equation} \label{3}
T=\frac{1}{2 \pi} \frac{\hbar a}{k_B c}.
\end{equation}
The second law of Newton $F= m a$ follows from substituting in (\ref{2}) equations (\ref{1}) and (\ref{3}).
Now is supposed that the boundary is a closed surface, it is assumed that is an sphere. Assuming that the holographic principle holds, the maximal storage space, or the total number of bits, is proportional to the area of the boundary
\begin{equation}\label{4}
N= \frac{A c^3}{G \hbar}= \frac{ 4 \pi R^2 c^3}{G \hbar},
\end{equation}
where a new constant $G$ is introduced. The total energy is given by the equipartition rule
\begin{equation}\label{5}
E= \frac{1}{2} N k_B T.
\end{equation}
Now we consider the total energy enclosed by the screen is given by a mass $M$ i.e. is satisfied $E=Mc^2$. Now equating this equation
with equation (\ref{5}) and substituting equations (\ref{4}) and (\ref{1}) we obtain the Newton's law of gravitation
\begin{equation}\label{6}
F= G \, \frac{m M}{R^2},
\end{equation}
and the constant $G$ is the universal gravitational constant. From this arguments it is stated in \cite{V} the entropic origin of gravity because the acceleration is related with an entropy gradient. More precisely, gravity
is explained as an entropic force caused by changes in the information associated
with the positions of material bodies. The consequences of this general theory
are being analyzed and discussed. The cosmological acceleration can be explained using the
entropic force, see \cite{EFS}. Other important consequence related to this work is that
the Verlinde's holographic model in an asymptotically de Sitter space leads to a new
form of the second law of motion which is the required by the MOND theory proposed
by Milgrom, see \cite{Fu}. Therefore the phenomenological Milgrom formulation is supported by Verlinde's theory.
In \cite{Fu} it is demonstrate that, in a
universe endowed by a positive cosmological constant $\Lambda$, the holographic model described
by Verlinde leads naturally to a modification of the second Newton's law of the form
\begin{equation}\label{fu1}
m [(a^2+k^2)^{1/2} -k] = F,
\end{equation}
where $k=\sqrt{\Lambda/3}$. Moreover equation (\ref{fu1}) is identical to the specific
formulation of MOND suggested by Milgrom in \cite{Mi5}. In the limit $a/k$ arbitrarily large (\ref{fu1})
becomes identical to the Newton second law and for $a/k \ll 1$ we have
\[
m \frac{a^2}{2k}= F,
\]
where $2k$ plays the role of the constant acceleration $a_0$. In fact, if
we assume that the present evolution of the universe is dominated
by the cosmological constant $\Lambda$, as corroborated by
observation \cite{TZH}, we can set $c H_0 \thicksim \Lambda^{1/2}$ which
implies that $k \thicksim a_0$ in orders of magnitude. The relation between
$a_0$ and the cosmological constant it is also discussed in \cite{G4} in the
context of the scaling laws that suggest a fractal universe.

\section{The Pioneer anomaly in this context}

The Pioneer anomaly \cite{A,A1} consists of unexpected, almost constant and uniform acceleration
 directed approximately towards the Sun $8.74 \pm 1.33 \times 10^{-10}
\ m s^{-2}$ first detected in the analyzed data of the Pioneer probes after they passed
the threshold of 20 Astronomical units. However, the recent new data of the Pioneer anomaly
suggest that it is variable and environment dependent rather than
a fixed value and still is not clear its direction with the possibility that be Earth directed, see \cite{TT2,TT}.
The effects of the Pioneer anomaly are non-detected on the major bodies of the solar system and
in several papers is studied its gravitational origin, see \cite{I5,T} and references therein.
Meanwhile there exits other works where it is studied its non-gravitational origin, see \cite{BFGP,RLLBD}.

The Pioneer anomaly is similar to the galaxy rotation problem which also
involves an unexplained acceleration. Milgrom realized
that MOND could explain the Pioneer anomaly, see \cite{Mi4}. The modified-inertia
approaches to solve the Pioneer anomaly have been also considered under
Unruh radiation by McCulloch, see \cite{Mc}. This proposal acquires its meaning
in the wake of the holographic scenario established in the work of Verlinde \cite{V}.
In \cite{Mc} it is found that the acceleration of the Pioneer craft is
given by
\[
a=\frac{G M_\odot}{r^2}+ \frac{\beta \pi^2 c^2}{\Theta},
\]
where $M_\odot$ is the sun mass, $\beta$ appear in the Wien's
constant and has the value $\beta=0.2$, and $\Theta$ is the Hubble
diameter $\Theta=2c/H_0=2 R_U$. The second term can be rearranged to
give
\begin{equation}\label{ess}
a=\frac{G M_\odot}{r^2}+ \frac{1}{2} \beta \pi^2 c H_0 \sim
\frac{G M_\odot}{r^2} + 0.99 \times cH_0.
\end{equation}
We are going to see that we obtain equation (\ref{ess}) in the
context of phenomenological formulation of MOND. We use equation
(\ref{for}) for the Pioneer craft, with the approximation $a_e
\sim a_0$ because we are dealing with a local object and taking
into account that the accelerations are quasi-collinear because
the Pioneer craft performs an orbit away from us (hence we can use
equation (\ref{eq})). In a strong Newtonian regime, we can develop the term
\begin{equation}\label{des}
\frac{a}{a+a_0}= \frac{1}{1+a_0/a} \approx 1- \frac{a_0}{a},
\end{equation}
in the case $a_0/a \ll 1$ i.e. $a_0 \ll a$. Now substituting in
(\ref{eq}) (taking the modulus) we have
\[
m_i \left(1- \frac{a_0}{a} \right) a=F= \frac{G M_\odot m_g}{r^2}.
\]
We can rearrange this equation to obtain
\begin{equation}\label{des2}
m_i \ a = \frac{G M_\odot m_g}{r^2} + m_i \ a_0 .
\end{equation}
From the equivalence principle we have $m_i=m_g$ and (\ref{des2}) becomes
\begin{equation}\label{eq2}
a= \frac{G M_\odot}{r^2}+\ a_0.
\end{equation}
In \cite{G} and \cite{G3} it is justified by different arguments
that $a_0 \sim c H_0$, where $H_0$ is the actual value of the
Hubble constant, see also \cite{G2}. Therefore we have obtained equation (\ref{ess}).
The arguments to obtain $a_0 \sim c H_0$ are the following. In \cite{G} using the equivalence principle, which
implies the equality  between inertial mass $m_i$ and gravitational mass $m_g$,
it is obtained the relation $GM_U = c^2R_U$ where $M_U$ and $R_U$
is the mass and the radius of the universe respectively. Then substituting
this expression in the definition of $a_0$ in the sense of the Mach's principle
the result follows. In \cite{G3} the relation $a_0 \sim c H_0$ is obtained through the scale factor
of the universe $R(t)$ and the Hubble law of expansion of the universe.
Hence, the Pioneer anomaly is given by $a_0$ that
taking into account that $H_0= 2.3 \pm 0.9 \times 10^{-18} s^{-1}$
we obtain that $a_0= 6.9 \pm 3.5 \times 10^{-10} \ m \ s^{-2}$,
which is in agreement with the observed value anomaly $8.74 \pm
1.33 \times 10^{-10} \ m s^{-2}$. The $40 \% \ (\pm 3.5)$
uncertainty arises because of uncertainties in the Hubble
constant.

\section{Final comments}

The value of $a_0=6.9 \pm 3.5 \times 10^{-10} \ m \ s^{-2}$ is
about six times larger than the acceleration constant $1.2 \times
10^{-10} \ m \ s^{-2}$ required for MOND of Milgrom for fitting
galaxy velocity curves. However, the constant acceleration $a_0$ is
also present in the inner solar system where it is dramatically
inconsistent with the
motion of the inner planets if we use equation (\ref{eq}) or
similar versions of $\mu(x)$. In fact, it fails completely in the
strong gravity regime where $a \gg a_0$, and thus cannot be valid
in the Solar system. For instance, the upper limit on an
additional constant acceleration imposed by the observed
precession of the orbit of Mercury is more than a factor of 10
smaller than $a_0$. Similar constrains result from the observed
motion of Icarus. In general, such peculiar acceleration is
constrained by observations to be about one or two orders of magnitude
lower than $a_0$ in the inner Solar system, see \cite{S}. Hence,
one must argue that the MOND acts in a very different way for local
bound objects like planets. This has already pointed out
by Milgrom in \cite{Mi4}. Sanders in \cite{S} concludes that
if the effects of the MONDian modification of gravity are not observed in the
motion of the outer planets in the solar system, the acceleration cannot be due to MOND.
Solar system constraints on multifield
theories of modified dynamics.
The $\mu(x)$ function (\ref{for}) used in this paper also present these problems.
Nevertheless, the result obtained for the Pioneer anomaly reinforces
that it must be Earth directed and variable and environment
dependent, because it depends on the relative position of the
Pioneer craft and the Earth in its own movement along its orbit
around the Sun.

A simple modification of the $\mu(x)$ function
does not save MOND from its inherent problems. In a recent review of the Pioneer anomaly
is said that the Pioneer anomaly has nothing to do with MOND, see \cite{TT}. In this
survey it is also said that ``the exact form of $\mu(x)$ remains unspecified
in both MOND and the relativistic version of TeVeS proposed by Bekenstein \cite{Be}.
It is conceivable that an appropriately chosen $\mu(x)$ might reproduce
the Pioneer anomaly even as the theory's main result, its ability to account for
galaxy rotation curves, is not affected". This also happens with the new expression
of $\mu(x)$ presented in this paper. It is still open to find the form of $\mu(x)$ consistent with
the observational data which establish differences between the
unbounded orbits (like the Pioneer craft) and the bounded orbits
(like the planets). In the framework of MOND, the internal dynamics of a
gravitating system $s$ embedded in a larger one $S$ is affected
by the external background field E of $S$ even if it is constant
and uniform, thus implying a violation of the strong
equivalence principle: it is the so-called External Field Effect (EFE).
Milgrom \cite{Mi1} originally introduced EFE in order to explain  that
the observed mass in certain open star clusters in the galactic neighborhood
of the solar system was very low, although their
internal accelerations were 5 or 10 times smaller than $a_0$. The galactic acceleration felt by
such open clusters is just of the order of $a_0$. The first, preliminary attempts to look at EFE
in the Oort cloud were made by Milgrom in \cite{Mi1,Mi11}. More detailed analysis on EFE in the Oort cloud is made
by Iorio in \cite{I6}.  EFE was adapted to the planetary regions of the Solar System,
where the field is strong, see \cite{Mi6}. Some implications were discussed in \cite{BN,I8,I9}.
Finally it should be mentioned that several studies of MOND were performed in the solar system,
see \cite{BMa,BN,I33,I44,I6,I8,I9,Mi5,S,SJ,TBHS}.
Anyway, the correct version of MOND to be constructed in the future must be derived
from the new holographic scenario. In \cite{H1,H2}, based on the hypothesis of the gravitational repulsion between
matter and antimatter, what allows considering, the
virtual particle--antiparticle pairs in the physical vacuum,
as gravitational dipoles, it is argued that the Pioneer Anomaly and the MOND is
related to the quantum vacuum fluctuations. Two speculative but exciting papers
which may help provide insight into the nature of the dark
energy of the Universe.

\acknowledgments
The author is partially supported by a MCYT/FEDER grant number
MTM2008-00694 and by a CIRIT grant number 2009SGR 381



\end{document}